\documentclass[varenna]{cimento}

\providecommand{\sorthelp}[1]{}

\newcommand*\bang{!}
\newcommand{\mytilde}{\raise.17ex\hbox{$\scriptstyle\mathtt{\sim}$}}

\def\spose#1{\hbox to 0pt{#1\hss}}
\def\simlt{\mathrel{\spose{\lower 3pt\hbox{$\mathchar"218$}}
     \raise 2.0pt\hbox{$\mathchar"13C$}}}
\def\simgt{\mathrel{\spose{\lower 3pt\hbox{$\mathchar"218$}}
     \raise 2.0pt\hbox{$\mathchar"13E$}}}

\usepackage{graphicx}  % got figures? uncomment this
\title{The Standard Model of Cosmology: A Skeptic's Guide}

\author{Douglas Scott\thanks{dscott@phas.ubc.ca}}

\institute{Dept.\ of Physics \& Astronomy, Univ.\ of British Columbia,
 Vancouver, Canada}

\begin{document}

\maketitle

\begin{abstract}
The status of the
standard cosmological model, also known as ``$\Lambda$CDM'' is described.
With some simple assumptions, this model fits a wide range of data, with
just six (or seven) free parameters.  One should be skeptical about this
claim, since it implies that we now have an astonishingly good picture of the
statistical properties of the large-scale Universe.  However, the successes
of the model cannot be denied, including more than $1000\,\sigma$ worth of
detection of CMB anisotropy power.  The model is older than most modern
astrophysicists seem to appreciate, and has not fundamentally changed for
more than a quarter of a century.  Tensions and anomalies are often discussed,
and while we should of course be open to the possibility of new physics, we
should also be skeptical of the importance of 2--3$\,\sigma$ differences
between data sets until they become more significant.  Still, today's SMC is
surely not the full story and we should be looking for extensions or
new ingredients to the model, guided throughout by a skeptical outlook.
\end{abstract}

\section{What is the standard model of cosmology?}
\label{sec:SMC}

The currently best-fitting picture for describing the statistics of the
Universe on large scales, the standard model of cosmology (or SMC),
is often known as $\Lambda$CDM, since it's a model in which the
matter is mostly cold and dark (i.e.\ effectively collisionless and with no
electromagnetic interactions, CDM), with the bulk of the energy density
of the Universe behaving like vacuum energy (i.e.\ like the cosmological
constant of general relativity, $\Lambda$).  But things are even more
specific than that, with the values of only about half-a-dozen free
parameters being enough to make a Universe that looks statistically just
like the one we live in -- and several of those parameters are now known to a
really impressive level of precision.  So the ``SMC'' is now quite
precisely prescribed.

It is an astonishing achievement of modern cosmology that we have come
to have such a successful model, especially when one considers that there
is no a posteriori reason to expect things to be this simple.
In physics we are driven to accept a model for several reasons -- certainly
that it fits the data, but also because of some less well defined notion
of {\ae}sthetics.  The simple group theoretical underpinnings of the
standard model of particle physics and the elegance of the field equations of
general relativity are obvious examples of this.  Sure, they fit lots and
lots of experimental data, but they're also really {\it nice\/}!
But for cosmology, no one would claim that the SMC is beautiful,
or even that it has to be correct because all alternatives
are uglier.  Certainly the SMC has some degree of simplicity (since it
doesn't need many free parameters), but why do those parameters have the
values that they do (see Sects.~\ref{sec:parameters} and \ref{sec:mnemonics})?
And why aren't there lots of other parameters required
(see Sect.~\ref{sec:information})?
Despite the fact that nothing in the basic cosmological picture has changed
since the early-to-mid-1990s (see Sect.~\ref{sec:oldness}),
most cosmologists are expecting something else to be just around the corner.
After all, surely the SMC can't be all there is?

In these notes I'd like to bring some attention to the idea that we should be
skeptical \cite{skepnote}
here, since we're dealing with very large themes.  A model that purports to
describe the whole of the observable Universe should be met with a decent
dose of incredulity!  It's important that we retain
a healthy level of skepticism when discussing any such claims.  But at the
same time we should also remember to be skeptical about {\it counter}-claims
(Sects.~\ref{sec:tensions} and \ref{sec:anomalies})
that haven't passed the same level of scrutiny.  And we should keep in mind
criteria that define what skepticism is (Sect.~\ref{sec:skepticism}), so that
we can isolate the successes of the SMC, while remembering that parts of
modern cosmology's lore remain quite speculative
(Sect.~\ref{sec:fundamentals}).

\section{The parameters and assumptions of the SMC}
\label{sec:parameters}

Let's be explicit about the standard model by giving the modern values of
its basic parameters.  Right now the determination of these quantities is
driven by cosmic microwave background (CMB) anisotropy experiments, and in
particular by results from the {\it Planck\/} satellite \cite{planck2014-a01}
(supported by many other kinds of data) on the CMB power spectra (which are
discussed further in Sect.~\ref{sec:information}).
Because of this, the basic parameter set is currently given in terms of the
quantities that are most directly measured by CMB experiments.  This means that
the parameters most often discussed in relation to observational constraints
are not necessarily the ones that are simplest to explain to the general
public, or that are the focus of non-CMB cosmologists.  These parameters are
listed in Table~\ref{tab:params}.  The set consists of: two densities,
$\Omega_{\rm b}h^2$ and $\Omega_{\rm c}h^2$ (for
baryonic matter and cold dark matter separately, since they have distinct
effects on the CMB power spectra), including a scaling of physical density
with the dimensionless
Hubble parameter, $h\equiv H_0/100\,{\rm km}\,{\rm s}^{-1}\,{\rm Mpc}^{-1}$;
a parameter $\theta_\ast$
that corresponds to the sound horizon divided by the angular diameter distance
to last scattering, which quantifies sliding the CMB power spectra left and
right; the amplitude $A_{\rm s}$ of the initial power spectrum of density
perturbations, defined at a particular scale, and often given as a logarithm;
the slope $n$ of the initial power spectrum as a function of wavenumber; and
a parameter $\tau$ describing how much the primary CMB anisotropies
are scattered by the reionised medium at low redshifts.

\begin{table}[htb!]
  \caption{Basic cosmological parameters, from a combination of {\it Planck\/}
  2015 data and other constraints from BAO, SNe and $H_0$ data
  (see Ref.~\cite{planck2014-a15}).  The CMB temperature comes from an analysis
  of the monopole spectral data by Fixsen \cite{Fixsen2009}.}
  \label{tab:params}
  \begin{tabular}{lcc}
    \hline
      Physical baryon density& $\Omega_{\rm b}h^2$& $0.02227\pm0.00020$\\
      Physical CDM density& $\Omega_{\rm c}h^2$& $0.1184\pm0.0012$\\
      Angular parameter& $100\theta_{\ast}$& $1.04106\pm0.00041$\\
      Reionisation optical depth& $\tau$& $0.067\pm0.013$\\
      Power spectrum amplitude& $\ln(10^{10}A_{\rm s})$& $3.064\pm0.024$\\
      Power spectrum slope& $n_{\rm s}$& $0.9681\pm0.0044$\\
    \hline
      CMB temperature& $T_0$ [K]& $2.7255\pm0.0006$\\
    \hline
  \end{tabular}
\end{table}

By far the best determined of these parameters is $\theta_\ast$, with a
signal-to-noise ratio ($S/N$) of about 2500 (from Table~\ref{tab:params},
or about 2300 from the CMB alone).  Then follows $A_{\rm s}$,
$\Omega_{\rm b}h^2$ and $\Omega_{\rm c}h^2$, with $S/N\simeq100$,
while $n_{\rm s}$ and $\tau$ only differ from their default values (of 1 and
0, respectively) at $S/N\simeq5$.  Other cosmological parameters that
are often discussed include $H_0$, $t_0$, $\Omega_{\rm m}$, $\Omega_\Lambda$,
$z_{\rm reion}$, etc., which are not independent, but
can be determined from the six parameters in the
context of the SMC.

Although it is often stated that there are six basic parameters, there's a
seventh that is often ignored.  This is the temperature of the CMB today
(or equivalently the radiation density), which is constrained using data from
the {\it COBE}-FIRAS instrument \cite{Fixsen1996}, as well as from several
other experiments (see Ref.~\cite{Fixsen2009}).  The determination is
systematics dominated, with $S/N\simeq5000$.  It is hence more precise
than other parameters, and dramatically better determined than other
densities.  For that reason it is usually considered to be fixed, and not
a free parameter at all.  However, the precision is starting to approach the
cosmic-variance limit, and so if $T_0$ was measured with much smaller errors,
we'd have to consider the fact that we can only measure parameters within
our Hubble patch and not actually ``background'' parameter values
(see Ref.~\cite{Zibin2008} for discussion).

But (to be skeptical about this), we might wonder whether there are other
hidden parameters.  There definitely are, to some extent, but mostly any
additions to the SMC are better cast as assumptions.  In fact
there are many of these, and it is important to be clear that the six (or
seven) parameters of the SMC are only descriptive of the Universe within a
specific framework.  A list of these assumptions is given in
Table~\ref{tab:assum} (and the reader can probably think of more).

\begin{table}[htb!]
  \caption{Some assumptions of the SMC.  Note that several of these apply
to our observable volume (which is the only part of the Universe that we
can test) only.}
  \label{tab:assum}
  \begin{tabular}{l}
    \hline
      \qquad Understanding the Cosmos is possible for human beings\\
      \qquad Physics is the same everywhere and at all times\\
      \qquad General relativity is the correct theory of gravity on
        cosmological scales\\
      \qquad The Universe is approximately statistically homogeneous and
        isotropic\\
      \qquad The Universe is spatially flat on large scales\\
      \qquad The dark energy behaves like a cosmological constant,
        with $w=-1$\\
      \qquad The dark matter is collisionless and cold for the purposes of
        cosmology\\
      \qquad There are three species of nearly massless neutrinos\\
      \qquad There are no additional light particles contributing to the
        background\\
      \qquad Density perturbations are adiabatic in nature\\
      \qquad The initial conditions were Gaussian\\
      \qquad The running of the primordial power spectrum is negligible\\
      \qquad The contribution of gravitational waves is negligible\\
      \qquad Topological defects were unimportant for structure formation\\
      \qquad The physics of recombination is fully understood\\
      \qquad One parameter is sufficient to describe the effects of
        reionisation\\
    \hline
  \end{tabular}
\end{table}

All of these assumptions are testable, and they all {\it have\/}
been investigated.
Many of them are tested through putting limits on extensions to the SMC,
e.g.\ checking whether the curvature is consistent with flat space, whether
there's evidence for modified gravity, non-trivial dark energy
(i.e.\ $w\,{\neq}\,-1$), or non-Gaussianity, or
whether there are signs of the effects of massive neutrinos or cosmic strings
(e.g.\ see Refs.~\cite{planck2013-p11,planck2014-a15}).

Nevertheless,
this is definitely a place where we need to exercise caution.  The confidence
with which we know the values of the basic set of six (or seven) parameters
depends on this being the full parameters space.  If there are more ingredients
in the actual model, then the parameters in the basic set will have
larger uncertainties.  For example, if we consider models that allow curvature
then the constraints on $w$ are very much weakened.  Hence we need to look
carefully at these tests.  Right now there is no strong evidence for any
additional parameter, but we fully expect that there will be
more ingredients required as the data improve, e.g.\ that the effects of
massive neutrinos or primordial gravitational waves
will eventually be measured.  And there may be genuine
surprises of course, like multiple kinds of dark matter or dark energy, or
important extra components, such as magnetic fields or isocurvature modes.

Nevertheless, there {\it has\/} been caution exercised, and despite attempts
to find evidence for additional parameters, the basic set continues to fit
very high signal-to-noise data -- particularly the CMB power spectra.

\section{The numbers that describe the Universe}
\label{sec:mnemonics}

Since it appears that the set of numbers required to statistically describe
the cosmological model has just seven elements, then these values become
important quantities that should be better known, among astronomers and
non-astronomers alike.
Many people follow the detailed statistics of their favourite sports teams,
or can name the capital cities of various countries, or give the sequence of
colours of the rainbow, or list the wives of Henry~VIII in order,
or name the actors who have played their favourite time-travelling alien.
Almost everyone learns the list of planets in the Solar System, through the
mnemonic about pizzas (that no longer includes pizza!).
So why don't most humans know the numbers that
describe the Universe that we live in?

Perhaps one of the problems is that the usual six parameters coming from
CMB anisotropies are quite esoteric.  This becomes apparent as soon as one
tries to explain the values in Table~\ref{tab:params} to the general public.
However, these six arcane numbers
(together with the assumptions that we've already discussed) span the space
of all parameters, and hence it's easy to present versions that are
simpler (like the age of the Universe, $t_0$, or the density of some
component, like $\rho_{\rm M}$) in more familiar units.
Let us highlight a few variants of 
quantities that are useful in describing our Universe, in the hope that
some of them may catch on!  Further examples along these lines can
be found in the paper ``Cosmic Mnemonics'' \cite{mnemonics}.

\begin{table}[htb!]
  \caption{Variants on the numbers that describe our Universe.}
  \label{tab:mnemomics}
  \begin{tabular}{l}
    \hline
\qquad 
      Characteristic scale on the CMB sky, $\theta_\ast\simeq0.6^\circ$
        (think eclipse{\bang})\\
\qquad 
      Radius of observable Universe $\simeq400\,{\rm Ym}$\\
\qquad 
      Age of the Universe $t_0\simeq5\,{\rm trillion}\ {\rm days}
        \simeq 5\times2^{200}t_{\rm Pl}$\\
\qquad 
      Age of the Universe is triple the age of the Earth,
        $t_0\simeq3\,t_\oplus$\\
\qquad 
      $H_0\,t_0$ is slightly less than 1, and $H\,t$ will be unity in about
        1 billion years\\
\qquad 
      $H_0$ will asymptote to the value
        $56\,{\rm km}\,{\rm s}^{-1}\,{\rm Mpc}^{-1}$ in the far future\\
\qquad 
      Cosmological constant, $\Lambda\simeq10^{-35}\,{\rm s}^{-2}$
        (``ten square attohertz'')\\
\qquad 
      Critical density, $\rho_{\rm crit}$, corresponds to 5 proton masses
        per cubic metre\\
\qquad 
      Density ratios,
        $\Omega_{\rm c}/\Omega_{\rm b}\simeq2\,\Omega_\Lambda/\Omega_{\rm m}
        \simeq5.3$\\
\qquad 
      Density parameter for photons, $\Omega_\gamma\simeq\alpha^2$\\
\qquad 
      Variance of density in spheres is unity at about $9\,{\rm Mpc}$
        (no $h^{-1}$)\\
\qquad 
      Amplitude of position-space density perturbations on Hubble scale,
        $\sigma\simeq6\times10^{-6}$\\
\qquad 
      Temperature at last scattering epoch $T_{\rm CMB}\simeq3000\,$K
        (think M giant{\bang})\\
\qquad 
      Age at last-scattering epoch, $t_{\rm rec}\simeq370\,$kyr\\
\qquad 
      Age at reionisation, $t_{\rm reion}\simeq600\,$Myr\\
\qquad 
      Number of particles in observable Universe $\simeq\alpha^{-42}$\\
    \hline
  \end{tabular}
\end{table}

With enough effort, it's easy to find numerological coincidences.  One
should obviously be skeptical about claims of significance for such things
though!  For example, from the
table we see that the number of particles in the observable Universe (mostly
photons) is about $\alpha^{-42}$ (where $\alpha$ is the fine-structure
constant), and additionally in the standard
model, the Earth forms at a redshift corresponding to $z\,{=}\,0.42$.
These facts could be used to suggest a link with Douglas
Adams' universal answer.

\vspace{0.5cm}
\begin{figure}[htbp!]
\includegraphics[width=0.50\textwidth]{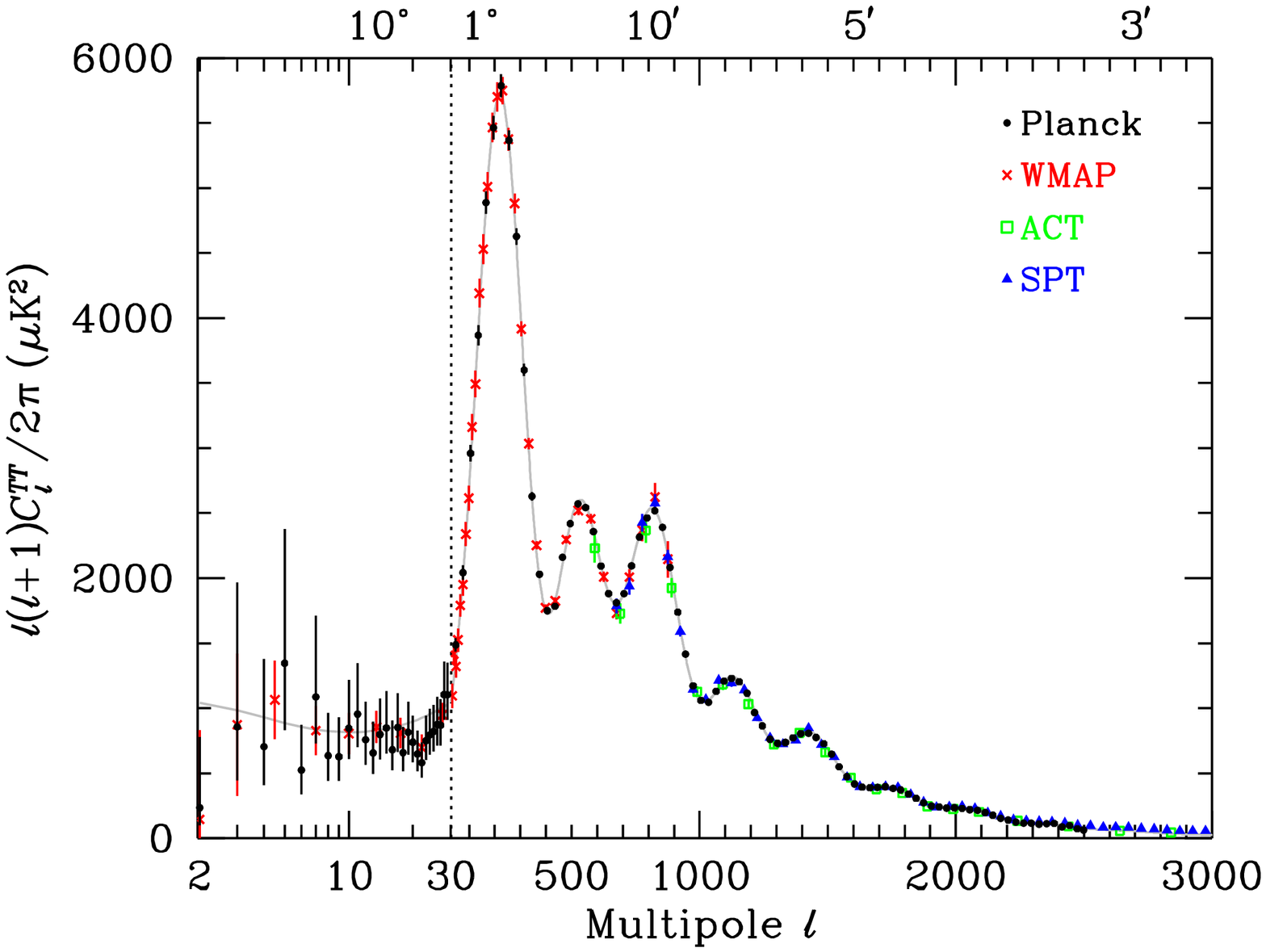}
\includegraphics[width=0.50\textwidth]{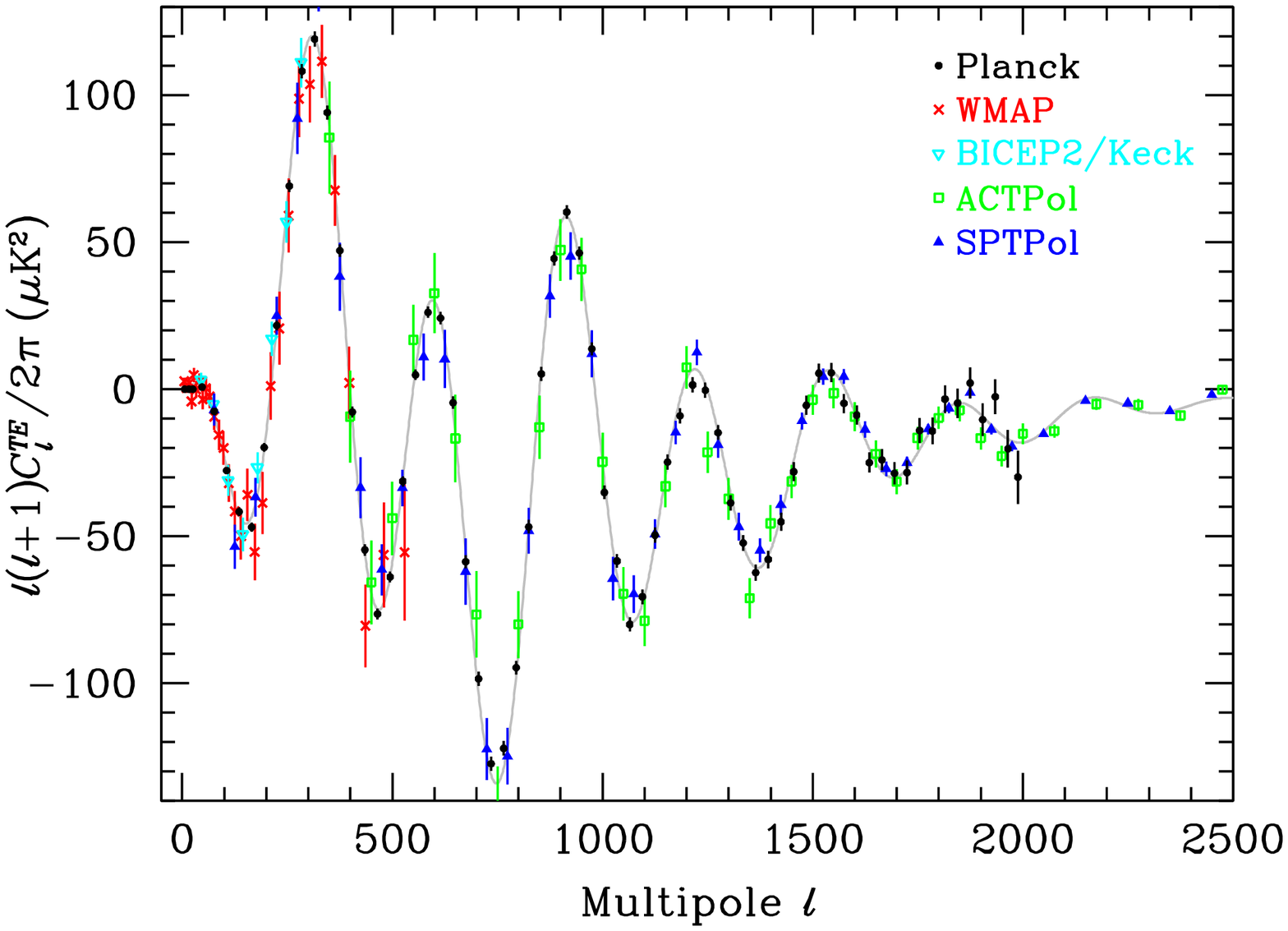}
\caption{CMB temperature anisotropy power spectrum (left) and
temperature-polarisation cross-power spectrum (right), from {\it Planck},
{\it WMAP}, BICEP/Keck, ACT and SPT (see Ref.~\cite{ScottSmoot2016} for full
references).  This demonstrates the current precision
with which these power spectra have been measured.}
\label{fig:TTandTE}
\end{figure}
\begin{figure}[htbp!]
\includegraphics[width=0.50\textwidth]{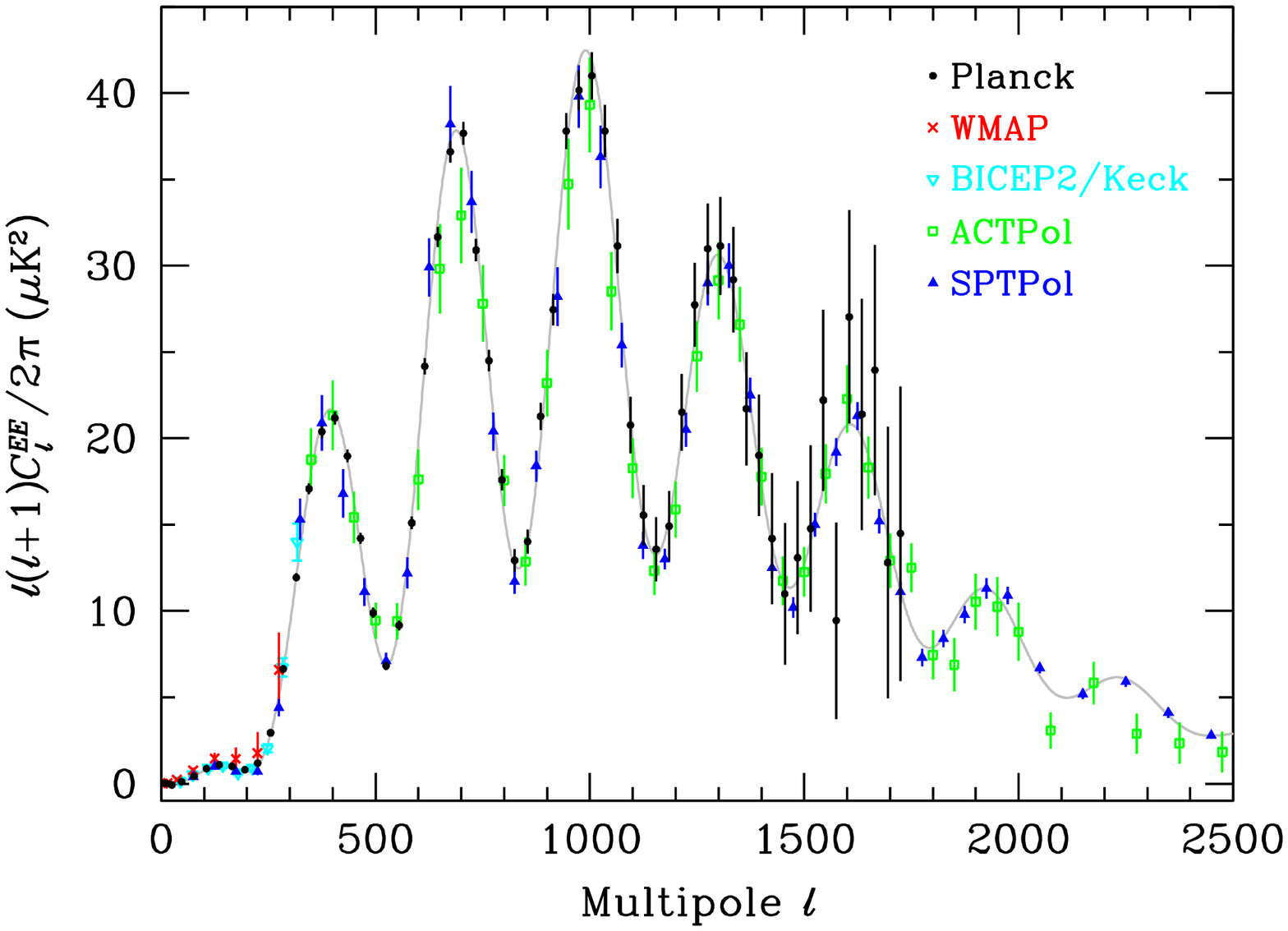}\
\includegraphics[width=0.50\textwidth]{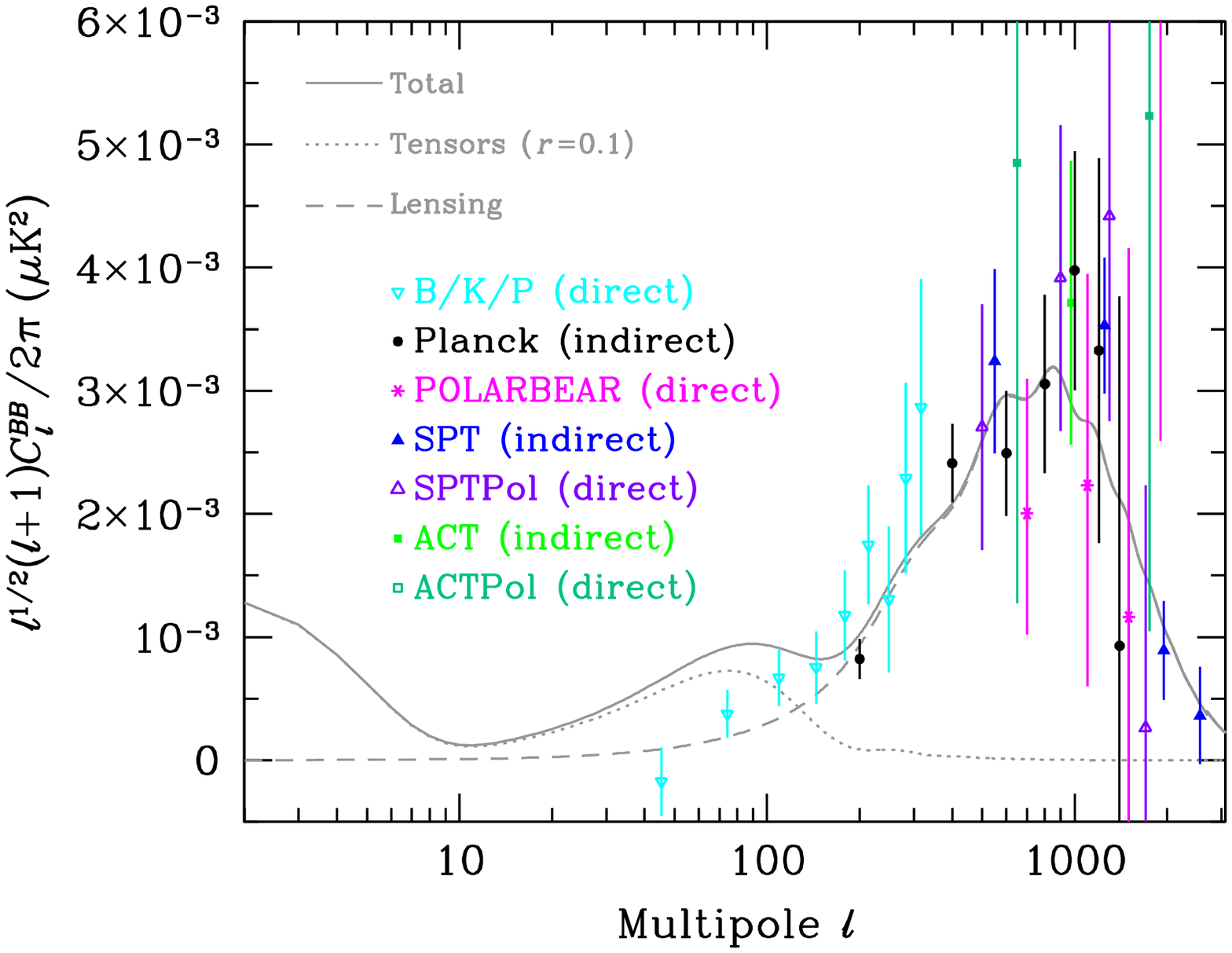}
\caption{Polarisation $EE$ (left) and $BB$ (right)
power spectra from several experiments (see Ref.~\cite{ScottSmoot2016} for
references).  The $BB$ spectrum here is scaled by a power of $\ell$ that makes
it possible to see all three of the expected peaks (from reionisation,
recombination and lensing).}
\label{fig:EEandBB}
\end{figure}

\section{Information in the SMC}
\label{sec:information}
Cosmological information comes from many sources.  However, at the present
time, the CMB dominates the constraints on the SMC.  This ``era of precision
cosmology'' can be seen through plots of the current status of the power
spectra coming from the CMB.  Figure~\ref{fig:TTandTE} shows the $TT$ (which
dominates the information) and $TE$ (which is catching up) spectra, while
Fig.~\ref{fig:EEandBB} shows the $EE$ (now also impressive) and $BB$ (still
in its infancy) spectra.  The grey line is the 6-parameter $\Lambda$CDM model
fit to the {\it Planck\/} $TT$ data, and one can see how well it matches the
other power spectra.

When we add up the total $(S/N)^2$ from the {\it Planck\/} CMB power
spectra, over the part of the sky conservatively believed to be free of
foreground emission, we find that the {\it Planck\/} $TT$, $TE$ and $EE$
measurements together correspond
to about $900\,\sigma$.  Adding the higher multipole measurements from
ACTPol and SPTPol means that today's CMB power spectrum determinations together
represent more than $1000\,\sigma$ of detection.  If we were skeptical about
the success of the SMC, then we should take note that it requires just
a few simplifying assumptions and seven free parameters to fit this huge
amount of information -- quite a remarkable achievement.

How does the constraining power of the CMB work?  The simple answer is that
it just depends on the number of modes that are measured, where the mode
amplitudes $a_{\ell m}$ come from expanding the sky as
$T(\theta,\phi)=\sum_{\ell m}a_{\ell m}Y_{\ell m}$.  Since the
anisotropies are Gaussian, then each $a_{\ell m}$ gives just a little bit
of information about the expectation value of the power in the $a_{\ell m}$s,
$C_\ell$ (or equivalently, the variance), and the total constraining power
is just about counting the number
of modes.  In more detail, if we have a cosmic-variance-limited experiment,
with $\Delta C_\ell =\sqrt{2/(2\ell+1)}C_\ell$, then the total
signal-to-noise ratio in the power spectrum is
\begin{equation}
(S/N)^2 \equiv \sum^{\ell_{\rm max}}_{\ell=2} (C_\ell/\Delta C_\ell)^2
 ={1\over2}\sum^{\ell_{\rm max}}_{\ell=2}(2\ell+1)
 ={1\over2}\left[\ell_{\rm max}(\ell_{\rm max}+2)-3\right]
 \simeq \ell_{\rm max}^2.
\label{eq:modes}
\end{equation}
But since the number of modes is just
$\sum_{\ell=2}^{\ell_{\rm max}}\sum_{m=-\ell}^{+\ell}$, then this means that
the total $(S/N)^2$ is just half the number of modes.

\begin{figure}[htbp!]
\includegraphics[width=\textwidth]{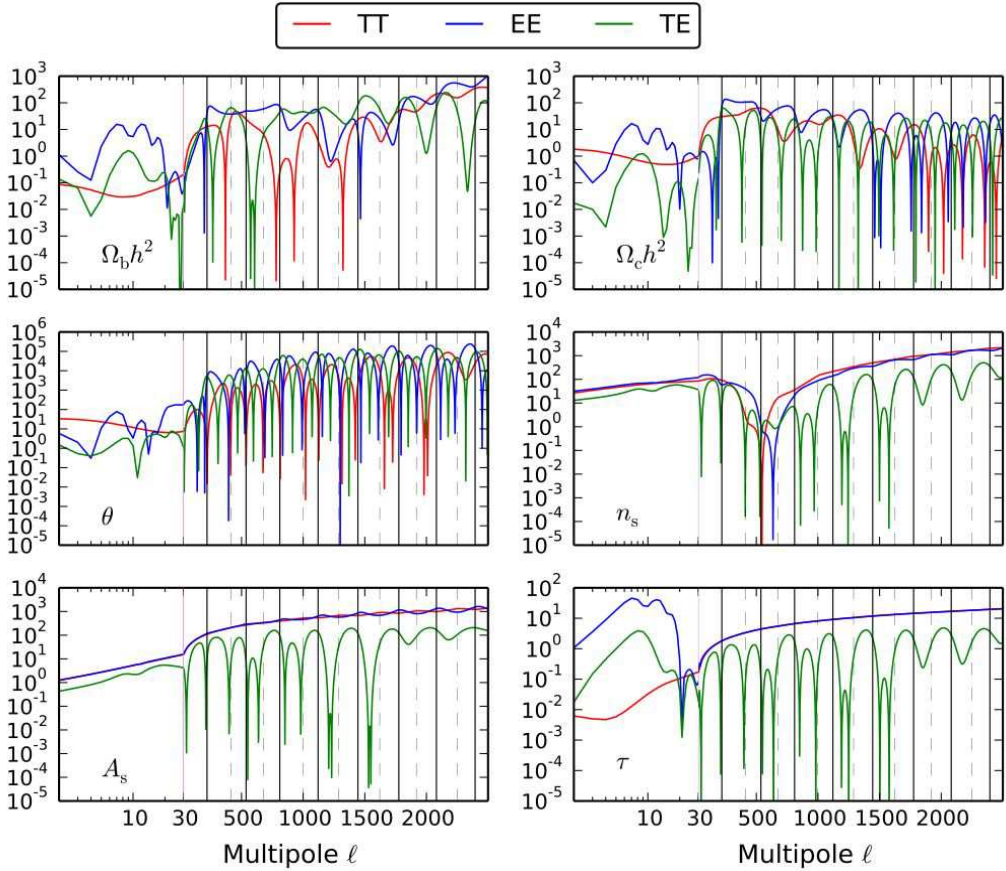}
\caption{Square of signal-to-noise ratio for each multipole, for a
cosmic-variance-limited experiment covering half the sky.  The results for
the $TT$, $TE$ and $EE$ power spectra are shown on panels for each of the
six parameters of the SMC, with a logarithmic horizontal axis for the first 30
multipoles.  Vertical lines mark peaks and troughs in the
$TT$ power spectrum.  These panels show the complex mapping of power spectrum
constraints onto parameters over different multipole ranges.}
\label{fig:SNR_versus_ell}
\end{figure}

To the extent that through {\it Planck}, we've measured all the modes out
to $\ell\simeq1500$ (over a large fraction of the sky), and the damping plus
foregrounds means that we can't go far beyond $\ell\simeq3000$ (say) for
primary CMB anisotropy measurements, then we're a good way through gathering
all the information we can get from $C_\ell^{TT}$.  But what about
polarisation?  Assuming for the moment that $C_\ell^{BB}$ is negligible, then
the existence of both $C_\ell^{TE}$ and $C_\ell^{EE}$ would seem to confuse
matters.  But really the situation is simply that we can measure the
scalar field $E$, in addition to $T$, for each pixel.  And hence, provided that
we measure both $C_\ell^{TE}$ and $C_\ell^{EE}$ out to some $\ell_{\rm max}$,
then we have exactly twice as much information as we would obtain from
$C_\ell^{TT}$ alone.  This means that the total ``information content''
can be defined to be just a count of the number of modes probed, in any of the
CMB fields.

Of course not all information is equal.  For example, {\it any\/} large-angle
$BB$ measurement would provide us with an entirely new kind of
information, enabling us to determine an additional parameter ($r$, the
tensor-to-scalar ratio) that is otherwise hard to constrain.  Moreover,
it is well known that adding polarisation data helps to break some
parameter degeneracies.  So we'd like to know how the parameter constraints
map onto the power spectrum modes.  The proper way to discuss this is through
the Fisher matrix, which includes derivatives of the power spectra with
respect to the cosmological parameters; this is demonstrated in
Fig.~\ref{fig:SNR_versus_ell} and discussed fully in
Ref.~\cite{ScottCNM16}.  One can see that some multipole ranges are
particularly important for some parameters, and as we go up in $\ell$, so that
new peaks or troughs are included, the constraining power can change
dramatically.

To focus on one example,
the behaviour of the $A_{\rm s}$ panel is simple -- if there
was only an overall normalisation to measure, then the constraints would
just come from the $S/N$ (and mode counting) expression given in
Eq.~(\ref{eq:modes}), with polarisation giving equal constraining power to
temperature.  In that sense, $A_{\rm s}$ is a ``linear'' parameter, since there
is a simple relationship between its total $S/N$ and the parameter constraint.
However, if the dependence is less trivial, then the relationship is
``non-linear'', and hence the way that the total $(S/N)^2$ is shared
out among the parameters is more complicated.  A good example is $\theta_\ast$,
which determines the amount by which the power spectra can be slid left and
right in multipole
-- this can be determined to great precision (because of the relative
sharpness of the acoustic peak structure), which is why this is the
best determined parameter of the SMC today.  In fact the total $S/N$ in
$\theta_\ast$ from {\it Planck\/} is around 2500, which is considerably more
than the {\it total\/} $S/N$ in the power spectra!

CMB polarisation has yet to become particularly constraining for the parameters
of the SMC.  But that situation will change as new experiments add modes,
doubling at least (or more, since high-$\ell$ $E$-mode measurements are not
as limited by foregrounds)
the information achievable from temperature alone, and providing
specifically useful degeneracy-breaking capability.
The discussion above can be extended to the $BB$ spectrum, as well as the
lensing spectrum $C_\ell^{\phi\phi}$ (which comes from the temperature
trispectrum, and gives an additional field, $\phi$).
This approach is useful for discussing future experimental constraints,
and how they map onto parameters.  But one thing it tells us is that
eventually we'll run out of CMB information.
This is essentially because the CMB information
is almost entirely restricted to two dimensions.
The same thinking can also inform discussions of more ambitious attempts to
extract the much larger amount of information contained in 3d surveys --
around $(c\,k_{\rm max}/H_0)^3$, if we can get all the $k$ modes down to scales
$k_{\rm max}$ \cite{MaScott16}.  This means that in principle we could one day
measure enough modes to give $\gg10^6\,\sigma$ of power detection.

\section{The venerableness of the SMC}
\label{sec:oldness}

It is clear that the standard model of cosmology is now well established.
So well
established in fact that a great deal of effort in modern cosmology is directed
towards trying to find extensions to the model.  For example, searching for
evidence that the dark energy is evolving or that more than two parameters 
are needed to characterise the perturbations.  Such searches for ``physics
beyond the standard model'' makes one think of similar endeavours to find
evidence to extend the Standard Model of particle physics.  When this sort
of thing comes up, it has been traditional for cosmologists to claim that
the SMC is relatively young and still in an exploratory stage --
so it's {\it nothing like\/} the chasing of
$3\,\sigma$ effects that appears to have motivated much experimental
particle physics for decades.  But in fact the SMC is actually quite long in
the tooth itself by now!

So how old is the SMC?  Certainly if one goes back a dozen years to a previous
overview by this same author \cite{SMC2006}, one finds little that has changed.
The model is much more precisely determined of course, but all the ingredients
are already in place.  Indeed, one can go back earlier, e.g.\ to the paper
``What Have We Already Learned from the Cosmic Microwave Background?''
(also known as ``What Has the CMB Ever Done For Us?'') written in 1998
\cite{CMBDoneForUs1999}, and find that the basic picture is just the same.
In fact many expositions of the history of cosmology state
that the model became established with the detection of cosmic acceleration
in 1998.  Of course that was an important part of the story, but I think it's
clear that what we now call $\Lambda$CDM was {\it already\/} the best-fitting
model when the supernova data came in and confirmed it -- to the extent that
even the most skeptical cosmologists had to take $\Lambda$ seriously.
Many papers had already pointed out before 1998 that a collection of results
pointed to a flat $\Lambda$ model being the best way of extending what had
previously been called ``standard CDM'' (or sCDM), i.e.\ a CDM-dominated model
with $\Omega_{\rm M}\,{=}\,1$ and an initial power spectrum of exactly the
Harrison-Zeldovich-Peebles form ($n\,{=}\,1$).  Among these results were:
the need for more power on large scales (to match galaxy clustering data);
the fact that most measurements of the density parameter tended to give
$\Omega\,{\simlt}\,0.3$; that few measurements of $H_0$ gave values in the
$\simeq50\,{\rm km}\,{\rm s}^{-1}\,{\rm Mpc}^{-1}$ range that were needed to
make the age of the Universe old enough for the stars within it; that the
amplitude of density perturbations from the {\it COBE\/} satellite's CMB
anisotropies pointed towards adiabatic perturbations in a $\Lambda$-dominated
(rather than open) model; and that indications from smaller-scale CMB
anisotropies were suggesting an acoustic scale consistent with flat geometry
\cite{ScottWhite94,PierpaoliSW00}.

Several of these arguments were compiled in two essays in 1995, one by
Ostriker \& Steinhardt \cite{OstSte95} and the other
by Krauss \& Turner \cite{KraTur95}.  Not everyone was
convinced of course, and some nostalgic theorists still tried to cling to
the Einstein-de Sitter elegance of sCDM \cite{WSSD95,WVLS96}.  However, the
writing was on the wall, and of all the flavours to add to sCDM, it was
apparent by the mid-90s that $\Lambda$CDM gave the best fit (even if you didn't
necessarily like it!).  Indeed it is possible to find earlier papers pointing
to this model being preferred by a combination of data -- and here the 1990
Nature paper by Efstathiou, Maddox \& Sutherland \cite{ESM90}
is a particular standout.  That's not to say that there weren't papers
proposing quite different models at the same time, but just that the currently
understood SMC was already there in the early 90s, with reasons to believe
that it provided the ``best-buy'' cosmology.

What this means is that the SMC is older than most people appreciate --
something like a quarter of century old, making it more than half the age
of the SM of particle physics!  As an indication of just how long ago that
was, in the early 90s we were using dial-up
modems to connect with the internet, the main browser was (the pre-Netscape)
Mosaic and the world's first text message was being sent!

\section{Tensions}
\label{sec:tensions}

The idea of ``tensions'' has already come up.  So let's take that particular
bull by the horns
right now.  There are several different minor chinks in the armour of
the SMC that are pointed to by various researchers.  A list of some of them
is given in Table~\ref{tab:tensions}.  None are sufficiently significant
to call them an actual discrepancy, hence the use of the word ``tension''
\cite{alternativenote}.  What should a skeptic make of an apparent disagreement
between different data sets at the 2--3$\,\sigma$ level?

\begin{table}[htb!]
  \caption{List of claimed tensions (not complete!).  Are any of these of
  consequence?}
  \label{tab:tensions}
  \begin{tabular}{l}
    \hline
      \qquad The amplitude $\sigma_8$ between CMB and cluster abundance\\
      \qquad Galaxy cosmic shear versus CMB constraints on
             $\sigma_8\Omega_{\rm m}^{0.5}$\\
      \qquad $H_0$ between traditional direct estimates and indirect
             CMB estimates\\
      \qquad {\it Planck\/} versus {\it WMAP\/} $TT$ power spectra\\
      \qquad {\it Planck\/} high-$\ell$ versus low-$\ell$ data\\
      \qquad Preference for $A_{\rm L}>1$ (apparent lensing effect) in
             {\it Planck\/} data\\
      \qquad Small-scale galaxy formation controversies\\
    \hline
  \end{tabular}
\end{table}

Well, let's remember that today's CMB data contains more than $1000\,\sigma$
worth of detection.  We can ask how many $3\,\sigma$ results there are in
$1000\,\sigma$.  Since signal-to-noise ratios add quadratically, the answer is
the number of times that $3^2$ goes into $1000^2$, and the answer is
{\it more than 100{,}000}.  So why is there so much focus on specific issues
that are barely at the $3\,\sigma$ level?

Of course part of the answer is that we shouldn't accept that the SMC is the
last word, but should keep an open mind to other possibilities.  There's also a
strong motivation to look for flaws that require revisions,
since we're all hoping to find fundamentally new physics by further confronting
the SMC with cosmological data.  And there {\it may\/} be evidence
of such things lurking in low $S/N$ disparities.  The trick of course is to
find the hints of disparity that grow from mere ``tensions'' into genuinely
significant differences.

However, concentrating on a few of the $>100{,}000$ potential 3-ish$\,\sigma$
effects seems like a misplaced kind of skepticism.  Trying to find minor
deficiencies in conventional wisdom seems to me to be a bit like chasing
conspiracy theories.  In any situation, you can always find {\it something\/}
that doesn't seem to make sense -- but you should be assessing the evidence
carefully, bearing in mind the context.  Here the context is: (1) that the
model (the SMC) fits a number of observational phenomena very well; (2) that
some of the uncertainties are of a systematic rather than statistical
nature; and (3) that there are a very large number of potential tensions
that could be selected from the $>1000\,\sigma$ of measured information.

\begin{figure}[htbp!]
\begin{center}
\includegraphics[width=0.75\textwidth]{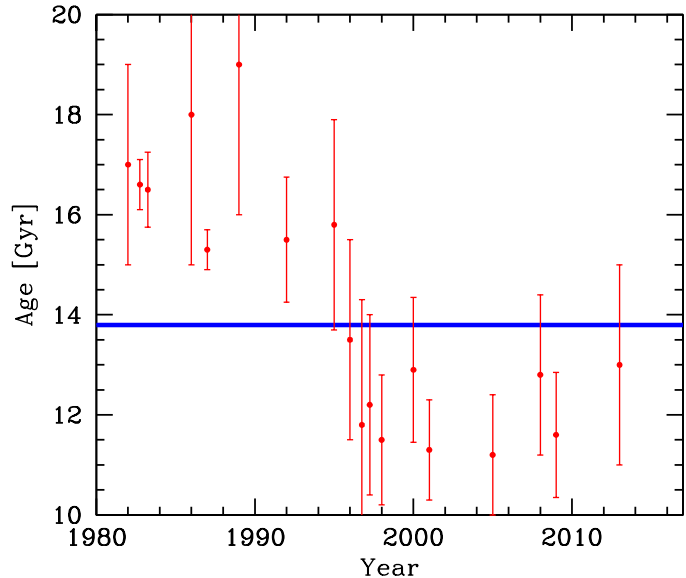}
\end{center}
\caption{Estimated ages of the oldest globular clusters, taken from a
representative set of review-like articles.  The preferred value dropped
dramatically in the mid-1990s.  The blue band is the current
best-fitting value for the age of the Universe, based on the 2015
{\it Planck\/} data (including uncertainty).}
\label{fig:cluster_ages}
\end{figure}

Let's look at a couple of aspects of the history of the development of the
SMC in order to see if there are any lessons we can learn.  Although the
SMC was already in place by the early 1990s, there are some observations
that have changed considerably since that time.  In particular, determinations
of the age of the Universe (from estimating the ages of the oldest
globular clusters) and determinations of the baryon abundance (coming from
Big Bang nucleosynthesis) changed in value in the mid-1990s.
Figure~\ref{fig:cluster_ages} shows the situation for cluster ages.  The
values plotted come from a representative selection of papers by several of
the groups working on this problem at that time.  In the early 90s the
oldest clusters were stated to be perhaps 17 billion years old, with lower
limits at around the 15 billion year level.  However, in hindsight it is
clear that those estimates were incorrect because they were dominated by
systematic uncertainties.  As other cosmological measurements improved, and
it became clear that the Universe had an age that was probably no more than
14 billion years, the cluster ages were revised to become consistent.  One
might imagine that the change could be traced to one particular effect that
was fixed -- but that really isn't the case.  Instead there were several
tweaks made over the years, most of which had the same sign and resulted in
the ages of the oldest clusters coming down to around 12 billion years.
The situation is really not very satisfying!  But I suspect this is often
the way things are when the uncertainties are to do with assumptions and
approximations in the analysis, rather than just being statistical.

The situation with baryon abundance is fairly similar.  But here it is harder
to make a plot of the values, since often there were no clear errors given!
Instead it was common to write down some feasible range for the
baryon-to-photon ratio, which was bracketed by different light-element
abundances (with little effort made at that time to designate 95\,\%
confidence ranges, or give $\pm1\,\sigma$ values, or the equivalent).
Despite this difficulty in interpreting the uncertainties in the old results,
what is clear is that the preferred value around 1990 was something like
$\Omega_{\rm b}h^2\simeq0.012$, while 10 years later it was around
0.022.  The change corresponded to a lot of $\sigma$ (with whatever
value of uncertainty you used).  Again, there wasn't one reason for this
change, but probably a list of things contributed to the increase (the
availability of damped Ly$\,\alpha$ systems for deuterium abundance
measurements being part of it).

\begin{figure}[htbp!]
\begin{center}
\includegraphics[width=0.75\textwidth]{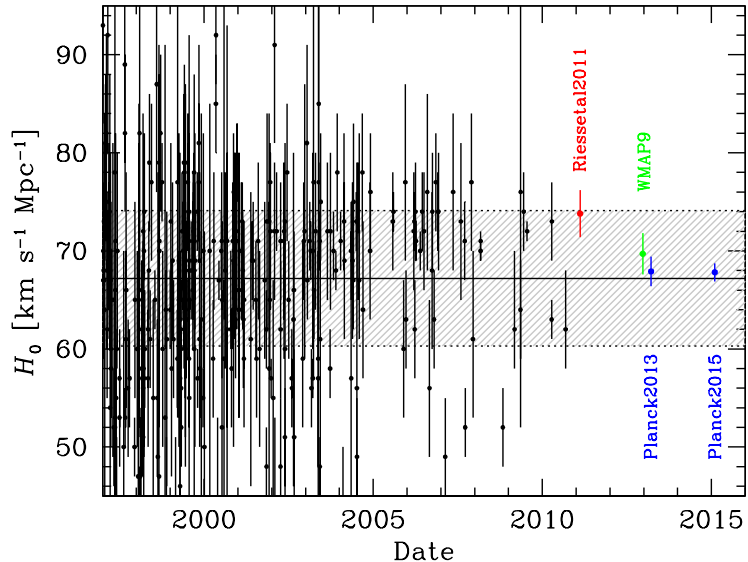}
\end{center}
\caption{Estimates of the Hubble constant between 1997 and 2010, compiled by
John Huchra (black points \cite{Hubbledata}).
These values were combined into a best estimate (using
practices for combining nuclear physics data) by Pritychenko \cite{NuclearH0},
with this value and uncertainty represented by the hatched region.  The new
value published by Riess et al.\ in 2011 \cite{Riess2011} is shown by the
red point.  The estimate derived from the full {\it WMAP\/} data set
is shown in green \cite{Bennett2013}.  The values obtained from the 2013
\cite{planck2013-p11} and 2015 \cite{planck2014-a15} {\it Planck\/} data
releases are the two blue points.}
\label{fig:hubble}
\end{figure}

The tension that perhaps attracts the most current attention is that some of
the most recent and precise values for the Hubble constant, determined using
standard distance-estimation techniques, appear higher than the value
determined from the SMC parameters that best fit CMB data.
Figure~\ref{fig:hubble} shows a large collection of published $H_0$
determinations from the period 1997--2010, including the error bars.  The
horizontal band gives an average of these data (using procedures developed
to deal with apparently disparate nuclear and particle physics data
\cite{NuclearH0}).  The newer value from Riess et al.\ (2011) \cite{Riess2011}
is indicated, together with values coming from {\it WMAP\/} and {\it Planck}.
Other recent values could be added, but wouldn't change the basic picture.
Placed in this context we can see the systematics-dominated history of
attempts to directly determine $H_0$ (which certainly goes back earlier than
1997!).  We have to assume that most sets of
authors of historical values believed that their published uncertainties were
a fair representation of their confidence in the data.  And yet it's clear
that at any given time the errors were being underestimated by most groups.
Perhaps the situation is genuinely different now, and the new (smaller)
error bars are correct.  But it's hard not to be skeptical.

In any case, we know that measurement techniques are continuing to improve,
and that if this CMB-versus-direct-determination tension in $H_0$ arises from
some genuinely new physics, then the statistical confidence in the
differences will grow.  Time will tell.

Turning to the example of {\it Planck}'s large-scale versus small-scale
constraints, there are some additional issues for the skeptic to keep in
mind.  As discussed in the Planck Collaboration Intermediate LI paper
\cite{planck2016-LI}, this situation is not as simple as it might at first
seem.  There are certainly parameter shifts when one considers {\it Planck\/}
low-$\ell$ versus {\it Planck\/} high-$\ell$ data, and these shifts may
seem to be at the 2--$3\,\sigma$ level.  However, the parameter space has six
(or maybe five, considering that $\tau$ is hardly measured) dimensions, with
many other directions in this space corresponding to particular parameter
combinations.  So one will see a 2$\,\sigma$ deviation in {\it some\/}
direction more than $5\,\%$ of the time, and hence it's necessary to take
into account the whole parameter set when assessing this kind of tension.
When that is done, the differences of low-$\ell$ versus high-$\ell$ parameters
have a probability to exceed above 10\,\% (i.e.\ nothing to write home about).
On top of that, it's unclear how one is choosing the angular scale to look
for a split.  And there is also a difficulty in assessing how unlikely it might
be for a data excursion to map onto a parameter shift.
The conclusion is that {\it Planck\/} and {\it WMAP\/} are in
spectacular agreement where they overlap, and the shifts seen at higher
multipoles are just about as big as you'd expect the shifts to be.
That's not to say that there might not also be problems with some of the
foreground modelling, or indeed some physics missing  from the SMC -- it's
just that the data don't {\it require\/} these things at the moment.

\section{Anomalies}
\label{sec:anomalies}

The other word that is much heard when discussing the CMB is ``anomalies''.
What is meant here is a feature (at a low level of significance) that appears
to be unexpected in the SMC, pointing to perhaps some kind of non-Gaussianity
or breaking of statistical isotropy.  There are several examples that have
been suggested over the years, with different researchers claiming
importance for one or other.  Table~\ref{tab:anomalies} gives a partial
list.  It seems extremely hard to believe that {\it all\/} of these are
pointing to deficiencies in the conventional picture.  One should be skeptical
of each of them, particularly because of the issue of a posteriori
statistics.  This issue is one that causes enough debate among cosmologists,
that I'm going to discuss it in some detail.

\begin{table}[htb!]
  \caption{List of claimed temperature anomalies.
  This is not intended to be a complete
  list (and several of these anomalies are related to each other).}
  \label{tab:anomalies}
  \begin{tabular}{l}
    \hline
      \qquad Low quadrupole and other low-$\ell$ modes\\
      \qquad Deficit in power at $\ell\simeq20$--30\\
      \qquad Low variance\\
      \qquad Lack of correlation at large angular scales\\
      \qquad The Cold Spot\\
      \qquad Other features on the CMB sky\\
      \qquad Hemispheric asymmetry\\
      \qquad Dipole modulation\\
      \qquad Alignment of low-order multipoles\\
      \qquad Odd-even multipole asymmetry\\
      \qquad Other features in the power spectrum\\
    \hline
  \end{tabular}
\end{table}

The problem is that all of these ``anomalies'' had their statistical
significance assessed {\it after\/} they were discovered.  Hence, in order
to fairly determine how unlikely they are, it is necessary to consider other
anomalies that may have been discovered instead.  Statisticians call this
the ``multiplicity of tests'' issue, which I think is the most helpful way
to think about it.

\begin{figure}[!htpb]
\begin{center}
\includegraphics[width=0.575\textwidth]{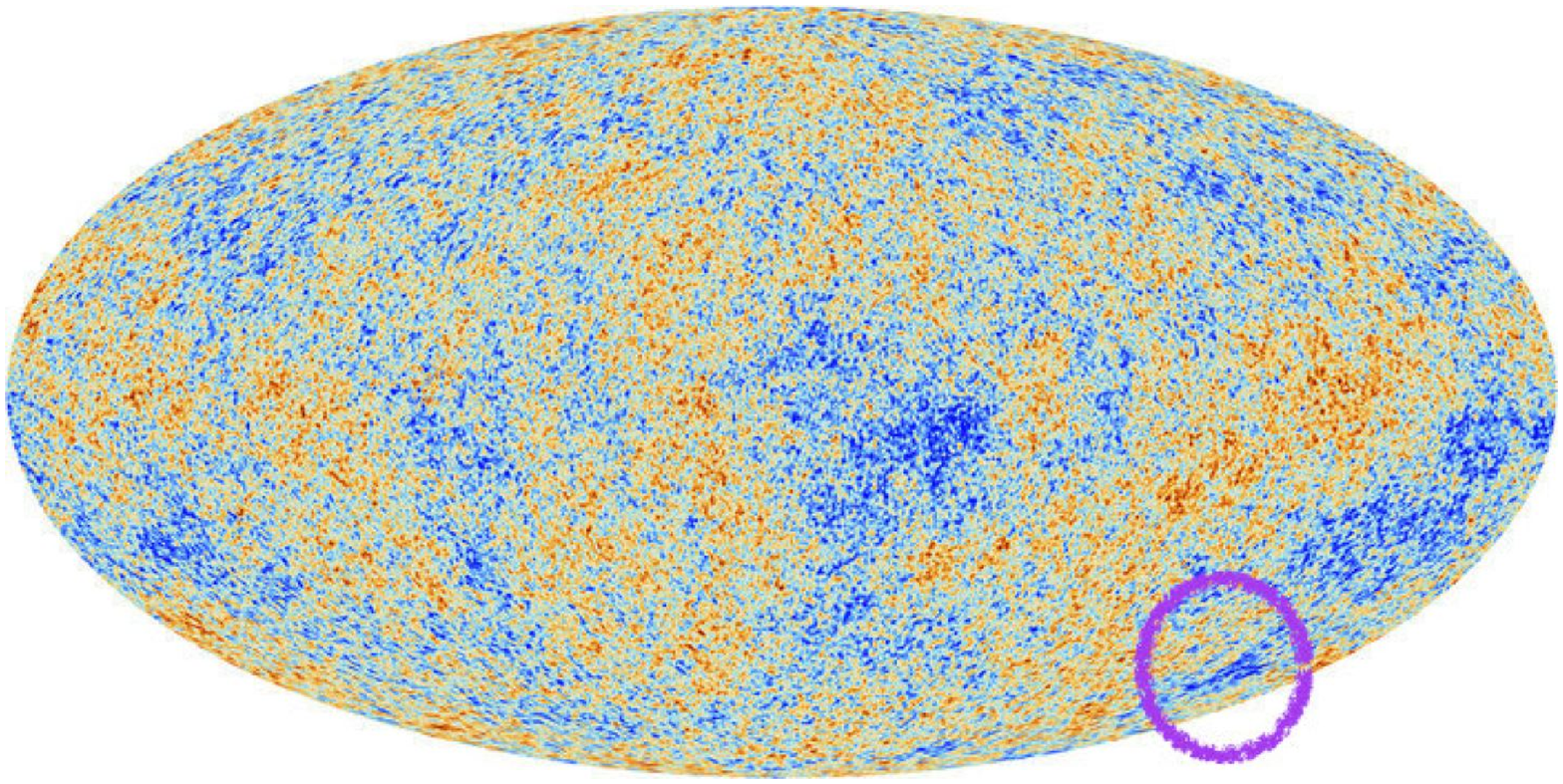}
% \hfill
\includegraphics[width=0.40\textwidth]{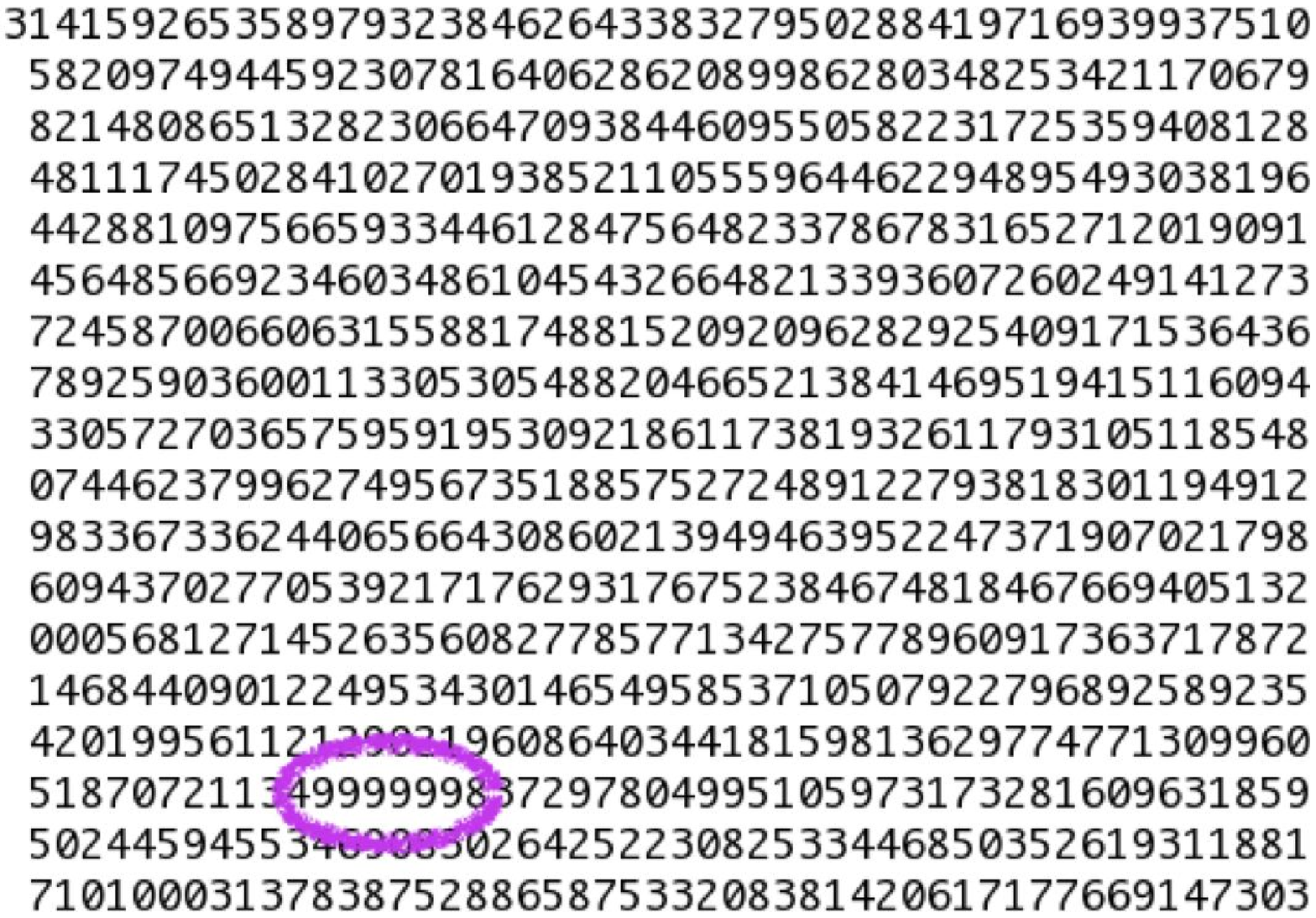}
\end{center}
\caption{\label{fig:nocoldspot}
Left: map of the CMB sky from {\it Planck}, with the position of the
so-called ``Cold Spot'' indicated.  Right: the first 900 digits of
$\pi$, showing the ``hot spot'' of six 9s (also known as the Feynman point).}
\label{fig:CSversusPi}
\end{figure}

Let's take the so-called CMB Cold Spot as an example (as shown in the left-hand
panel of Fig.~\ref{fig:CSversusPi}).  The probability of finding a cold region
of exactly this size and shape in exactly this particular direction is
obviously vanishingly small.  No one would consider such a calculation to be
useful, and
at the very least would appreciate that the spot could have been found in
any direction -- hence to assess the significance one could look in simulated
skies for similarly extreme
cold spots that occur anywhere.  The probability determined in this way
then becomes of order
$0.1\,\%$.  However, a specific scale was chosen for the spot, or to be
more explicit, a filter of a particular shape and scale was chosen.  It turns
out that the Cold Spot isn't very extreme if a purely Gaussian filter is
used, but is pulled out at higher amplitude by a ``compensated'' filter (e.g.\
what is often called a ``Mexican hat'') with a scale of about $5^\circ$.  This
means that one should marginalise over the scale (within some reasonable
bounds) and over a set of potential filter shapes (that one might have chosen)
as well.  On top of all this it's obviously clear that one needs to consider
{\it hot\/} spots as well as cold spots (this may seem so self-evident that
it doesn't need to be stated, but in fact several papers have {\it only\/}
assessed the significance of cold spots).  And the situation is more complicated
than that, since if there had been a fairly conspicuous pair of neighbouring
spots, or a hot spot diametrically opposite a cold spot, or even a triangle
of spots, then one might equally well have been writing papers about the
anomalous feature that was discovered.

The point is that in each Hubble patch (with the CMB sky being an independent
realisation of the underlying power spectrum), there will be features on the
sky or in the power spectrum, that appear anomalous.  One has to consider the
set of potential anomalies in each patch in order to assess whether a
feature is extreme enough to get excited about.  In practice 2--$3\,\sigma$
anomalies go away when you marginalise over these possibilities, but
${\simgt}\,5\,\sigma$ anomalies would remain anomalous after maginalisation.

A criticism of this way of thinking is that it's just {\it too\/}
skeptical!  The argument is that if you try hard enough to marginalise over
possible tests then you can make {\it anything\/} appear to be insignificant.
I don't think this is true, since you have to be reasonable here (as in any
assessment of statistical evidence, where there is always some
subjectivity).  And I stand
by the claim that it's hard to make $5\,\sigma$ effects go away, while
2--$3\,\sigma$ effects that are subject to a posteriori statistics should
always be viewed with extreme skepticism.

\vskip 0.3cm

Another way to look at this is to make an analogous study of something that
you are confident is genuinely random.  This was done in the paper with
Dr.\ Frolop \cite{PiInTheSky}, comparing the CMB anomalies with patterns in the
digits of $\pi$.  Several examples are given there, but let's just pick one.
As illustrated in the right panel of Fig.~\ref{fig:CSversusPi}, there are six
consecutive occurrences of the digit ``9'' at the 762nd digit of $\pi$.
Assuming that the digits are random, a simple calculation (considering the
number of ways of placing the run of 9s in the first 762 digits, and the
number of ways of picking the other 756 digits) gives a probability
of $756/10^6$.  Of course the run of six numbers didn't have to be 9 (even
although you could consider 9 to be special because it's the highest),
and hence the probability for any run of six
is 10 times greater -- but that's still less than $1\,\%$!  So why does this
not shake our faith in the digits of $\pi$ being random?  The answer is that
a posteriori statistical effects can be subtle.  In this particular case,
we would have found an equal probability for obtaining a run of {\it five\/}
9s at an earlier digit, or {\it four\/} 9s even earlier, and so on.  When
including this hidden ``multiplicity of tests'', the
probability becomes of order $10\,\%$, i.e.\ not small enough to decide
that there are messages written in the digits of $\pi$!  And in addition to
all of that, there are other patterns that might also have been remarked upon
if they had been found,
a run of ``123456'' for example, or an alternating series like ``9090909''.
Perhaps these aren't quite {\it as\/} striking as six 9s, but they should get
{\it some\/} weight in considering the set of tests, and hence in the
assessment of the significance of the anomaly.  I'm convinced that if
one went to the trouble of looking for such things, there would appear
to be something conspicuous
in most chunks of $\pi$ (in every 1000 digits, say) -- just like there are
some apparent ``anomalies'' in every Hubble patch's CMB sky.

Despite all these words of caution, let me add that
one should {\it still\/} continue the search for anomalies, since any genuinely
significant large-scale oddities could be signs of exciting new physics
(e.g.\ see discussion in Ref.~\cite{CHMSZ}).
And of course sometimes $3\,\sigma$ things will become $5\,\sigma$ when
more data are included -- so it's worth continuing these investigations.
A problem is that the large-angle temperature field
has already been well mapped, and those data are now limited by cosmic
variance.  So the only way to make progress is to include new data, such as
from CMB polarisation \cite{CZSBG}.  What would be particularly good would
be to find some kind of natural explanation for an anomaly, with no (or very
few) free parameters, which also makes a clear prediction for some {\it new\/}
observable, such as polarisation, lensing, or 21-cm observations.
If there's such a prediction, and a $3\,\sigma$ result is found, then that
really would mean 99.7\,\% confidence.

\section{The nature of skepticism}
\label{sec:skepticism}

I've said a lot about being skeptical -- but what does that really mean?
What I'm talking about here is the concept philosophers might call ``scientific
skepticism'', which involves questioning assertions that lack empirical
evidence.  I believe this to be a fundamental part of scientific inquiry.
It can
be summed up through the phrase ``extraordinary claims require extraordinary
evidence'' (popularised by Carl Sagan) -- and obviously that applies well to
cosmology, through its grandiose themes, just as it does to pseudo-science.
Science isn't completely mechanical and dispassionate, since it
includes speculation and creativity as part of the process of development
-- but that's not the same as
accepting every new idea that comes along.  At the other end of the spectrum,
it's also important not to fall into the trap of ``denialism'', i.e.\
adopting a position that rejects every claim {\it even if\/} there's good
evidence to support it (like climate change, or dark matter perhaps).

To be a bit more explicit about skepticism,
let me pick the writings of a particular modern
philosopher, namely Mario Bunge, who has written extensively on the topic
of scientific epistemology.  Among other definitions, he describes
how any authentic science must include ``changeability,
compatibility with the bulk of the antecedent knowledge, partial intersection
with at least one other science, and control by the scientific community''
\cite{Bunge}.  These ideas give a little more content to the notions of
hypothesis testing, falsifiability, parsimony, etc., that we learn
about in school.  And they make clear that the skeptical approach is central
to the establishment and evolution of scientific ideas.

As examples of topics that fail to meet these criteria and land up in the
pseudo-science
category, Bunge lists ``astrology, alchemy, parapsychology, characterology,
graphology, creation `science', `intelligent design', Christian
`Science', dowsing, homeopathy, and memetics''.  However, Bunge also states
that ``cosmology is still rife with speculations that contradict solid
principles of physics''!  He says that for good reason -- the SMC lives within
the domain of ``physical cosmology'' and has passed a wide array of tests,
but, on the other hand, the most theoretical aspects of cosmology are indeed
in an entirely different conjectural realm.  Hence it is important to separate
the concrete parts of modern cosmology (the answers to the ``what'' questions)
from the areas where we are still speculating wildly (and trying to find
answers to the ``why'' questions).

\section{Beyond the SMC}
\label{sec:fundamentals}

We'd all like to understand where the whole Universe comes from, or explain
away the dark matter and dark energy.  Speculation is certainly good,
but {\it believing\/} your speculation (before it has
passed any tests) is bad science!  The correct approach should be to 
investigate the consequences of your idea and try to determine if there are
definitive predictions that can be confronted with data.

I feel that there's a kind of malady that infects some cosmologists, where
pretty much any outlandish and unorthodox idea is considered at the same
level as the conventional picture -- rather than giving it a higher degree of
skepticism, like all extraordinary claims deserve.
Perhaps part of the blame here is that modern physics in general, and the
SMC in particular, contain some fairly bizarre-sounding concepts.  We teach
students about quantum mechanics and black holes, that we can build a model for
the whole observable Universe, that there are hypothetical particles that
dominate all matter and that a negative-pressure fluid is driving the cosmic
acceleration.  So perhaps students start to think that
any hare-brained scheme is equally worth pursuing?

I can't shake the feeling that a good dose of skepticism would help keep
things in perspective.

An example of this is inflation (see \cite{planck2013-p17} and
\cite{planck2014-a24} for discussions).  It is undoubtedly an appealing idea,
and there is a great deal of circumstantial evidence to support it -- so I
think it's entirely reasonable to be a {\it fan\/} of inflationary cosmology.
But since inflation is really a framework rather than a model, we can't
assert that any of the observations actually prove that inflation is correct
\cite{inflationnote}.  It seems reasonable to assume that whatever picture
turns out to describe the early Universe, and generates the perturbations, it
will contain some of the features of the current inflationary paradigm.  But
I don't think we can proclaim that we know that it will include {\it all\/} the
ingredients -- not until we have some more direct evidence.

However, one of the problems with assessing the merits of inflation is that
there isn't a good alternative.  Sure, there might be some ideas suggested
as counter-proposals, but they tend to seem much more ad hoc,
or create more problems than
they solve, or have predictions that are less well developed.
And the same issue applies more broadly across other ``alternative''
theories.  The SMC has been developed over decades and the calculations are
relatively straightforward (involving Gaussian perturbations, linear theory,
well-understood physics, etc.) -- but there's no reason to expect the same
to be true for some unconventional new idea.  So if an alternative is
proposed, then it's not trivial to determine whether it can match the
precision tests of the SMC.  We just have to be a little patient until the
calculations can be done accurately enough.

Despite the need to be open to alternatives, when there are clear
predictions, it's still important to be skeptical if they just don't fit
the data.  As an example, we
call the dominant form of energy in the Universe ``dark energy'', as though
its properties were mysterious and unknown -- and a huge amount of
effort is going into measuring its equation of state ($w$ as a function of
redshift) with increasing precision.  But the reality is that all measurements
so far are consistent with this component being simply vacuum energy with
$w=-1$.  I've
heard people say that it's much more likely to be a model with $w\neq-1$, since
$w=-1$ has zero probability!  But really, there's no sensible model that
gives a definite prediction other than pure vacuum,
and so we're left with the notion that there's
just a universal constant, $\Lambda$, that gives a small (but non-zero)
energy density to empty space.

Another example is dark matter.  It's obvious that an alternative
explanation for galaxy rotation curves {\it might be\/} that we can modify our
theory of gravity.  And there have been several suggestions along those lines
(see e.g.\ \cite{Fin}).
However, the evidence for dark matter comes from a lot more than rotation
curves of galaxies, e.g.\ the depth of cluster potential wells and
measurements of gravitational lensing.  But in fact the most robust evidence
for dark matter comes from the CMB anisotropies -- there is {\it no\/} model
for fitting the power spectra that doesn't include a lot more CDM than
baryonic matter.  Here we have a choice between abandoning GR (or even
Newtonian gravity) or just imagining that there's a component of matter that's
not very shiny!  Even without guidance from data, it seems fairly clear that
the parsimonious explanation is to have a particle that's like a heavier
version of the neutrino.  But the skeptic should come down more heavily on the
side of CDM when comparing with clustering, lensing and (particularly) CMB
data.

A related issue is the evaluation of some of the small-scale puzzles associated
with galaxies.  It has become common to propose models ascribing these to some
property of the dark matter (just strong enough to detect, without messing
up the SMC predictions entirely).  However, galaxy formation is a complicated
business \cite{EOGF}, involving non-linear complexity, hydrodynamics,
feedback processes, etc.  Since we {\it know\/} that we don't fully
understand baryonic physics, we should be skeptical of assertions that some
new property of dark matter has been discovered because of indications coming
from non-linear scales.

Despite the examples given here, the SMC is in no sense a complete model, and
there will surely be several additions eventually.  Table~\ref{tab:questions}
lists some potential questions relating to physics {\it beyond\/} the SMC.
Will one of these lead to the next breakthrough?
Right now the path to progress isn't at all clear.  Maybe it will turn out
to be something else entirely, something unexpected and outlandish -- but only
if the evidence strongly supports that.

\begin{table}[htb!]
  \caption{Physics beyond the SMC.  Which of these questions will turn out
  to be fruitful?}
  \label{tab:questions}
  \begin{tabular}{l}
    \hline
      \qquad Where did the parameters come from\?\\
      \qquad Did inflation happen\?\\
      \qquad Can we explain the value of $\Lambda$\?\\
      \qquad Why is $\Omega_{\rm c}/\Omega_{\rm b}\simeq5.3$\?\\
      \qquad Are any anomalies or tensions worthy of attention\?\\
      \qquad Can we detect primordial gravitational waves\?\\
      \qquad Can we detect primordial non-Gausianity\?\\
      \qquad Are there missing ingredients to the SMC\?\\
      \qquad Will neutrino properties be measurable\?\\
      \qquad Can we predict reionisation from first principles\?\\
    \hline
  \end{tabular}
\end{table}

\section{Conclusions}
\label{sec:conclusions}

I've given an overview of the current status of the Standard Model of
Cosmology, the SMC, and stressed how important it is to maintain a healthy
level of skepticism when assessing the successes of this model, and in
evaluating the merits of extensions to it.

So when should one be skeptical and when not?  That's the trick of course!
Obviously the aim is to be {\it right}, and it's never clear how to forecast
the future.  There was a time when hardly anyone believed that the solution
to the Solar neutrino problem lay in the properties of neutrinos -- but a small
number of people got it right before the rest of us.  Similarly, some people
saw that $\Lambda$CDM fit most of the data while many others in cosmology
were working on things like ``open CDM'' or ``mixed dark matter''.  Since
no practicing cosmologist believes that the current SMC will be the last word
on a statistical description of the Universe, then there are surely
developments that are yet to come.  The goal is (somehow) to pick the
2--$3\,\sigma$ effects that grow to be important parts of the model -- and
by implication, part of this process involves ignoring most of the other
claims for chinks in the SMC's armour.

There were times in the history of cosmology when it was fairly clear what
directions were going to be fruitful for pursuing calculations or
observations.  I think it's not just that we have the benefit of hindsight
-- it really was the case that at one time studying hot versus cold dark
matter was obviously a good idea, and at some other time developing the
theory of CMB anisotropies or building experiments to probe degree-scale
anisotropies were clearly worthwhile.  However, right now it's not at all
obvious where cosmology is going next.

This means that this is either the worst time or the best time to be a
cosmologist!  If you have a good idea (and it turns out to be right) you
could find yourself on your own making the next major contribution to our
understanding of the whole of the Cosmos.

\acknowledgments
I acknowledge many enjoyable discussions about some of the issues contained in
this contribution with members of the Planck Collaboration, including those
who were at UBC, particularly Dago Contreras, Ali Narimani and Jim Zibin.

\bibliographystyle{varenna}
\bibliography{varenna_dscott}

\begin{thebibliography}{10}
\bibitem{skepnote}
I'm going to use the American English spelling of this word, since to many
  non-British people, ``sceptic'' looks too much like ``septic''.
\bibitem{alternativenote}
I'm reminded of the skeptics' joke that when ``alternative medicine'' works
  it's just called ``medicine''. In the same way, if a ``tension'' was actually
  big enough to be significant, then it would be called a ``difference''.
\end{thebibliography}

\end{document}